# Frequency-dependent topological phases and photonic detouring in valley photonic crystals

Guo-Jing Tang[1,†], Xiao-Dong Chen[1,†], Fu-Long Shi[1], Jian-Wei Liu[1],
Min Chen[1,2,*], and Jian-Wen Dong[1,*]

1, School of Physics & State Key Laboratory of Optoelectronic Materials and Technologies, Sun Yat-sen University, Guangzhou 510275, China.

2, Department of Physics, College of Science, Shantou University, Shantou 510063, China.

[†]These authors contributed equally to this work

[*]corresponding author: stscm@mail.sysu.edu.cn & dongjwen@mail.sysu.edu.cn

## ABSTRACT

The recent exploration of the valley degree of freedom in photonic systems has enriched the topological phases of light and brought the robust transport of edge states around sharp bends. The two and more simultaneous band gaps in valley-Hall systems have attracted researchers' attention for enlarging the working bandwidth. However, band gaps with frequency-dependent topologies were not reported and the demonstrated flow of electromagnetic waves is limited to the robust transport of edge states. Here, the frequency degree of freedom is introduced into valley photonic crystals with dual band gaps. Based on the high-order plane wave expansion model, we derive an effective Hamiltonian which characterizes dual band gaps. Metallic valley photonic crystals are demonstrated as examples in which all four topological phases are found. At the domain walls between topologically distinct valley photonic crystals, frequency-dependent edge states are demonstrated and a broadband photonic detouring is proposed. Our findings provide the guidance for designing the frequency-dependent property of topological structures and show its potential applications in wavelength division multiplexers.

# I. Introduction

Topological photonics [1-4] has attracted much attention for its practical applications in photonic diodes, robust waveguides, high-Q cavities and low-threshold lasers [5-15]. Recently, the valley, which marks the energy extrema of band structure, has been widely studied in two-dimensional layered materials [16-18] and borrowed into the photonic realm [19-33]. Around two inequivalent Brillouin zone corners, i.e., K' and K points, bulk bands have opposite Berry curvature. As a result, the bulk band is characterized by a nonzero valley Chern number and there are deterministic edge states within the band gap above the bulk band. Up to now, as the time-reversal symmetry is preserved, valley-Hall topological phases and the resultant edge states can be achieved in all-dielectric photonic crystals [19, 28], helical waveguides [23], designer surface plasmon crystals [21], and plasmonic crystals [32, 33]. Many intriguing phenomena such as robust delay lines [19], perfect refraction [22], tunable excitation [28], reconfigurable transmission [31] have been proposed or demonstrated. On the other hand, the frequency degree of freedom is widely used in communication and information processing. The utilization of frequency degree of freedom makes it possible to transfer and process signals at multi-frequencies and increase the channel capacity of photonic devices [34-37]. Based on the frequency degree of freedom, many multiband devices such as multiband filters [38], multiband transceivers [39], multi-functional photonic crystal fiber splitters [40], are proposed.

Recently, two or more simultaneous band gaps are discussed in valley-Hall systems and they enlarge the working bandwidth of waveguides [41, 42]. However, band gaps with frequency-dependent topologies were not reported and the demonstrated flow of electromagnetic waves was limited to the robust transport of edge states. In this work, we introduce the frequency degree of freedom into valley photonic crystals (VPCs) with dual band gaps. To model dual band gaps at two different frequency ranges, an effective Hamiltonian based on the high-order plane wave expansion model is obtained. All four topological phases of dual band gaps are found by changing two structural parameters of the designed metallic valley photonic crystals. Frequency-dependent edge states are demonstrated at domain walls between VPCs with frequency-dependent topologies. With three joint VPCs, the broadband photonic detouring is demonstrated by switching the operating frequency. Our work may have potential applications in multiband photonic devices such as wavelength division multiplexers.

## II.    Effective Hamiltonian

Without loss of generality, we consider the two-dimensional triangular photonic crystals (PCs) [Fig. 1(a)]. We focus on the transverse magnetic modes, and hence the out-of-plane relative permittivity $\varepsilon_z$ and the in-plane relative permeability $\mu_\parallel$ are concerned. By considering the Maxwell equations with the nonzero electromagnetic fields of ($H_x$, $H_y$, $E_z$) and replacing $H_x$ and $H_y$ by $E_z$, we obtain:

$$\omega^2 \varepsilon_0 \mu_0 \varepsilon_z(\mathbf{r}) E_z(\mathbf{r}) + \left(\partial_x^2 + \partial_y^2\right) E_z(\mathbf{r}) = 0 \tag{1}$$

To protect the structural symmetry and achieve frequency isolated Dirac cones, we focus on PCs whose unit cell possesses $C_3$ rotation symmetric permittivity distribution $\varepsilon_z(\mathbf{r})$ [e.g., one presented in Fig. 1(a)]. By applying the periodicity of the PC, we can expand $E_z(\mathbf{r})$ and $\varepsilon_z(\mathbf{r})$ as $E_z(\mathbf{r}) = \sum_{\mathbf{G}} E_{\mathbf{G}} e^{i(\mathbf{q}+\mathbf{G})\cdot\mathbf{r}}$, $\varepsilon_z(\mathbf{r}) = \sum_{\mathbf{G}} \beta_{\mathbf{G}} e^{i\mathbf{G}\cdot\mathbf{r}}$ where $\mathbf{q}$ is the reciprocal wave vector and $\mathbf{G}$ is the reciprocal lattice vector. Substitute them into Eq. (1), we get equations for each $\mathbf{G}$:

$$\omega^2 \varepsilon_0 \mu_0 \sum_{\mathbf{G'}} (\beta_{\mathbf{G}-\mathbf{G'}} E_{\mathbf{G'}}) - \left[(q_x + G_x)^2 + (q_y + G_y)^2\right] E_{\mathbf{G}} = 0 \tag{2}$$

To analyze the Dirac cones locating around the corners of the Brillouin zone, we first consider K point and its adjacent $k$-points. This makes $q_x = K + \delta k_x$ and $q_y = \delta k_y$, where $K = \dfrac{4\pi}{3a}$. To analyze dual band gaps locating at two different frequency ranges, we consider the first set of three reciprocal lattice vectors which makes $|\mathbf{K}+\mathbf{G}_i|=K$, i.e., $\mathbf{G}_0=(0,0)$, $\mathbf{G}_1=(-3K/2,\sqrt{3}K/2)$, $\mathbf{G}_2=(-3K/2,\sqrt{3}K/2)$, and the second set which makes $|\mathbf{K}+\mathbf{G}_i|=2K$, i.e., $\mathbf{G}_3=(-3K,0)$, $\mathbf{G}_4=(0,\sqrt{3}K)$, $\mathbf{G}_5=(0,-\sqrt{3}K)$ [Fig. 1(b)]. By considering these six reciprocal lattice vectors and omitting two singular bands, the valley-Hall dual band gaps are described by the effective Hamiltonian [see details in Appendix A]:

$$\delta \hat{H} = \hat{v}_D \left( \hat{\tau}_z \hat{\zeta}_z \hat{\sigma}_x \delta k_x + \hat{\sigma}_y \delta k_y \right) + \lambda \hat{\sigma}_z \tag{3}$$

where ($\delta k_x$, $\delta k_y$) measures from the K or K' point, $\hat{v}_D$ is the photonic Dirac velocity. $\hat{\sigma}_i$, $\hat{\tau}_i$ and $\hat{\zeta}_i$ are the Pauli matrices acting on sub-lattice, valley spaces and band gap spaces, respectively. The last term $\lambda \hat{\sigma}_z$ opens a frequency band gap with the bandwidth of $2\lambda$. This effective Hamiltonian indicates that the valley Chern number is given by $C_{vn} = \hat{\zeta}_z \text{sgn}(\lambda_n)$. Here $\lambda_1 = i \frac{\sqrt{3}\varepsilon_0\mu_0\omega_{D1}^3}{4K^2}\left(\beta_{\mathbf{G}_2}-\beta_{\mathbf{G}_1}\right)$,

$\lambda_2 = i \frac{\sqrt{3}\varepsilon_0\mu_0\omega_{D2}^3}{16K^2}\left(\beta_{2\mathbf{G}_2}-\beta_{2\mathbf{G}_1}\right)$, where $\omega_{D1} = K\left[\varepsilon_0\mu_0\left(\beta_{\mathbf{G}_0}-\beta_{\mathbf{G}_1}/2-\beta_{\mathbf{G}_2}/2\right)\right]^{-1/2}$ and

$\omega_{D2} = 2K\left[\varepsilon_0\mu_0\left(\beta_{\mathbf{G}_0}-\beta_{2\mathbf{G}_1}/2-\beta_{2\mathbf{G}_2}/2\right)\right]^{-1/2}$. It indicates that the valley Chern numbers, i.e., $C_{v1}$ and $C_{v2}$, are determined by the signs of $\left(\beta_{\mathbf{G}_2}-\beta_{\mathbf{G}_1}\right)$ and $\left(\beta_{2\mathbf{G}_2}-\beta_{2\mathbf{G}_1}\right)$. As the valley Chern number is quantized to be -1 or 1, we can achieve four and only four topological phases of dual band gaps characterized by ($C_{v1}$, $C_{v2}$), and this can be realized by carefully designing the unit cell of PCs.

### III. Dual band gaps and topological phases

To realize all four topological phases, we consider the metallic VPC as a concrete example. As depicted in Fig. 1(a), the lattice constant is $a$ = 20 mm. The unit cell contains a circle-bar shaped metallic rod embedded in the background whose permittivity is $\varepsilon$ = 2.25. The metallic rod consists of a circle with the radius of $r$, six bars with the length of $l$ and the width of $w$. Throughout this work, we fix these three structural parameters to be $r = 6.57$ mm, $l = 7.53$ mm, and $w = 0.5$ mm. Six bars are classified into three same groups which are separated by an angular distance of 120°. The central angular position of each group is given by $\theta$ and the angular distance between two bars within each group is given by $2\alpha$. They are limited by -60° < $\theta$ ≤ 60° and 2.18° ≤ $\alpha$ ≤ 30° due to the $C_3$ rotation symmetry and the size of the metallic rod. Then, we have two structural parameters, i.e., ($\theta$, $\alpha$) to change the unit cell of VPCs and observe the topological phase transition. Figure 1(c) shows bulk bands of the VPC with $\theta$ = -10° and $\alpha$ = 8° (named as VPC1). From 10 to 22 GHz, there are dual band gaps whose frequencies respectively range from 12.50 to 13.26 GHz (i.e., Gap 1) and from 19.01 to 19.58 GHz (i.e., Gap 2). Although both dual band gaps are omnidirectional, they are characterized by different valley Chern numbers which can be inspected by the eigen-fields of bulk states just below the band gap [28]. To see this, $|E_z|$ fields and energy fluxes of the first (fourth) bulk state, which is just below Gap 1 (Gap 2), are shown in Fig. 1(d). The first bulk state has its $E_z$ fields vanishing and has clockwise energy flux vortex at triangular-lattice centers $p$ of the unit cell, and we label it as the $p^-$ state. On the contrary, the fourth bulk state has its fields vanishing and anti-clockwise energy flux vortex at triangular-lattice centers $q$ of the unit cell, and we label it as the $q^+$ state. Bulk states below Gap 1 and Gap 2 have contrasting field distributions. It leads that Gap 1 is characterized by the valley Chern number of $C_{v1} = 1$ and Gap 2 by $C_{v2} = -1$.

We can reconfigure the metallic VPCs by changing the structural parameters of $\theta$ and $\alpha$, to have different topological phases and the associated phase diagram [Fig. 2(a)]. Note that the diagram only shows the topologies of PCs with -30° ≤ $\theta$ ≤ 30° as those of PCs with $|\theta|$ > 30° can be predicted by the symmetry analysis. In the phase diagram, the dashed (solid) curve plots the boundary where the closing of both Gap 1 and Gap 2 (Gap 2) happens. These two curves divide the whole diagram into four domains which are indexed by Roman numerals from I to IV. In each domain, VPCs are characterized by a pair of valley Chern numbers, i.e. ($C_{v1}$, $C_{v2}$). For example, the above discussed VPC1 is characterized by (1, -1) and locates in domain I [marked by the red dot]. Phase transition can be observed by changing or $\theta$ and $\alpha$. For example, we first consider the case of keeping $\alpha$ but changing $\theta$. Figure 2(b) shows the evolution of bulk states at the K point as a function of $\theta$ (keeping $\alpha = 8°$). Around the phase transition point at $\theta = 0°$, the first and second bulk states come closer, meet each other, and move apart. Similar case is also found between the fourth and fifth bulk states around Gap 2. After the state exchange, both Gap 1 and Gap 2 experience the topological phase transition, and the resultant VPC is characterized by $C_{v1} = -1$ and $C_{v2} = 1$ and locates in domain II. As an example, we consider the VPC with $\theta = -10°$ and $\alpha = 8°$, e.g., the VPC2 [marked by the blue dot]. Figure 2(c) shows its corresponding bulk bands and the eigen-fields of bulk states below dual band gaps. The first bulk state has its fields vanishing and has anti-clockwise energy flux vortex at triangular-lattice centers $p$ of the unit cell, while the fourth bulk state has its fields vanishing and has clockwise energy flux vortex at triangular-lattice centers $q$ of the unit cell [Fig. 2(c)]. The difference between these field distributions is in agreement with the fact that dual band gaps have opposite valley Chern numbers.

Within domain I and domain II, $C_{v1}$ and $C_{v2}$ are opposite. However, there are another two topological phases in which $C_{v1}$ and $C_{v2}$ are the same, i.e., phases in domain III and domain IV [Fig. 2(a)]. To realize these new phases, we consider the case of changing VPC1 by keeping $\theta$ but changing $\alpha$. Figure 2(d) shows the evolution of bulk states at the K point as a function of $\alpha$ (keeping $\theta = -10°$). There is no exchange between the first and the second bulk states and hence the topology of Gap 1 keeps unchanged. On the contrary, there is an exchange between the fourth and the fifth bulk states, indicating the change of the topology of Gap 2. As an example, we consider the VPC with $\theta = -10°$ and $\alpha = 20°$, named as VPC3 and marked by the green dot in Fig. 2(a). Its bulk bands are shown in Fig. 2(e) along with eigen-fields of two bulk states. As both fields are vanished and has clockwise energy vortex at triangular-lattice centers $p$, the first and fourth bulk states are both $p^-$ states, indicating Gap 1 and Gap 2 have the same topology. This is in agreement with the fact that $C_{v1} = C_{v2} = 1$. Based on the symmetry analysis, the topological phase with $C_{v1} = -1$ and $C_{v2} = -1$ can be achieved in the VPC with $\theta = 10°$ and $\alpha = 20°$. So far, all four frequency-dependent topological phases have been demonstrated in our proposed VPCs.

## IV. Frequency-dependent edge states and broadband photonic detouring

With the obtained four topological phases, we can realize frequency-dependent edge states. To see this, we first consider the domain wall between VPC1 at the top and VPC2 at the bottom [inset of Fig. 3(a)]. For VPC1 and VPC2, Gap 1 (Gap 2) is respectively characterized by $C_{v1} = 1$ and $C_{v1} = -1$ ($C_{v2} = -1$ and $C_{v2} = 1$). Therefore, these two VPCs are topologically distinct and edge states should be found in both dual band gaps. Figure 3(a) shows the dispersion of the corresponding edge states [marked in red]. There are edge states covering the whole dual band gaps. In addition, the differences of valley index at the K valley crossing the domain wall are -1 and 1 for Gap 1 and Gap 2. As a result, the direction of the group velocity of edge states within Gap 1 at the K valley is negative, while that within Gap 2 is switched to be positive. Figures 3(d) and 3(e) show $E_z$ fields of two representative edge states at the K valley at 9.19 GHz in Gap 1 and 16.11 GHz in Gap 2. The Fourier transforms of the field distributions of both edge states are presented and both have high intensity around the K points. For edge state in Gap 1, the highest $k$-components locate within the first Brillouin zone [right panel of Fig.

3(d)]. But for edge states in Gap 2, some high-order *k*-components are excited [Fig. 3(e)]. This is reasonable as we should expand with high-order plane waves to analyze eigen-states with higher frequency. In Fig. 3(a), we achieve valley-dependent edge states in both dual band gaps. On the other hand, we can selectively control the existence of edge states in Gap 1 or Gap 2 with the obtained topological phases. For example, we consider the domain wall between VPC1 at the top and VPC3 at the bottom [inset of Fig. 3(b)]. As two VPCs have Gap 1 with the same topology, there is no edge states in Gap 1 which is confirmed by the numerical results [Fig. 3(b)]. But their Gap 2 share different topology, i.e., $C_{v2}$ = -1 for VPC1 and $C_{v2}$ = 1 for VPC3, so edge states are found in Gap 2. As a result, edge states are designed to exist in Gap 2 but not in Gap 1. We can also design domain wall for having edge states existing in Gap 1 but not in Gap 2. This is achieved at the domain wall between VPC3 at the top and VPC2 at the bottom [Fig. 3(c)]. As these two VPCs have Gap 1 with different topologies but Gap 2 with the same topology, there are edge states in Gap 1 but not in Gap 2. By employing the frequency degree of freedom, we can not only control the propagating direction of edge states, but also control the existence of edge states in dual band gaps, i.e., frequency-dependent edge states can be realized.

With the presented frequency-dependent edge states, we can realize frequency multiplexing flow of light. For example, we can achieve the broadband photonic detouring by switching the operating frequency [Fig. 4]. The proposed photonic device is schematically shown in Fig. 4(a). The left part of the photonic device is the domain wall between VPC1 and VPC2, while VPC3 is inserted into them at the right part. The source is input from the port 1 at the left and received from port 2 or port 3 at the right. According to edge dispersions shown in Fig. 3, the channel from port 1 to port 2 only supports edges states in Gap 1 while the channel from port 1 to port 3 only supports edge states in Gap 2. Hence, by switching the operating frequency from Gap 1 to Gap 2, the flow of light is detoured from the lower channel to the upper channel. This is confirmed by the numerical results presented in Figs. 4(b) and 4(c). When edge states in Gap 1 (e.g. at 12.95 GHz) is excited, they are guided to the port 2 [Fig. 4(b)]. When the operating frequency is switched to be in Gap 2 (e.g. at 19.10 GHz), the output flow of light is redirected to port 3. To check the performance of the proposed photonic detouring, we calculate the transmission from port 1 to port 2 and port 3 (i.e. $S_{21}$ and $S_{31}$) as a function of frequency [Figs. 4(d) and 4(e)]. $S_{21}$ is dominant when the frequency is operating in Gap 1, i.e., 12.80 GHz to 13.26 GHz. However, $S_{31}$ is dominant when the frequency is operating in Gap 2, i.e., 19.01 GHz to 19.39 GHz. The broadband high transmission is inherent from the wide frequency ranges of edge states. Hence, with four topological phases and the associated frequency-dependent edge states, we can achieve the photonic detouring which has potential applications in wavelength division multiplexers.

## V. Conclusion

In conclusion, we propose the high-order plane wave expansion model and analyze the frequency-dependent topological properties in dual band gaps. With the proposed circle-bar shaped metallic VPCs, we realize dual band gaps and the topological phase transition by changing the angular parameters $\theta$ and $\alpha$. All four frequency-dependent topological phases are demonstrated in a topological phase diagram. Frequency-dependent edge states are found at the domain walls and it leads to the broadband photonic detouring by switching the operating frequency. The introduction of frequency degree of freedom into valley-Hall photonic crystals makes it flexibly mold the flow of light. For instance in optical communication, the photonic detouring has potential applications in wavelength division multiplexers. The frequency- and valley-dependent edge states also have other potential applications involving frequency related operations.

## Acknowledgements

This work is supported by National Natural Science Foundation of China (Grant Nos. 11761161002, 61775243, and 11704422), Guangdong Basic and Applied Basic Research Foundation (Grant Nos. 2018B030308005 and 2019B151502036), State Key Research Development Program of China (Grant No. 2019YFB2203502), and Science and Technology Program of Guangzhou (Grant No. 201804020029).

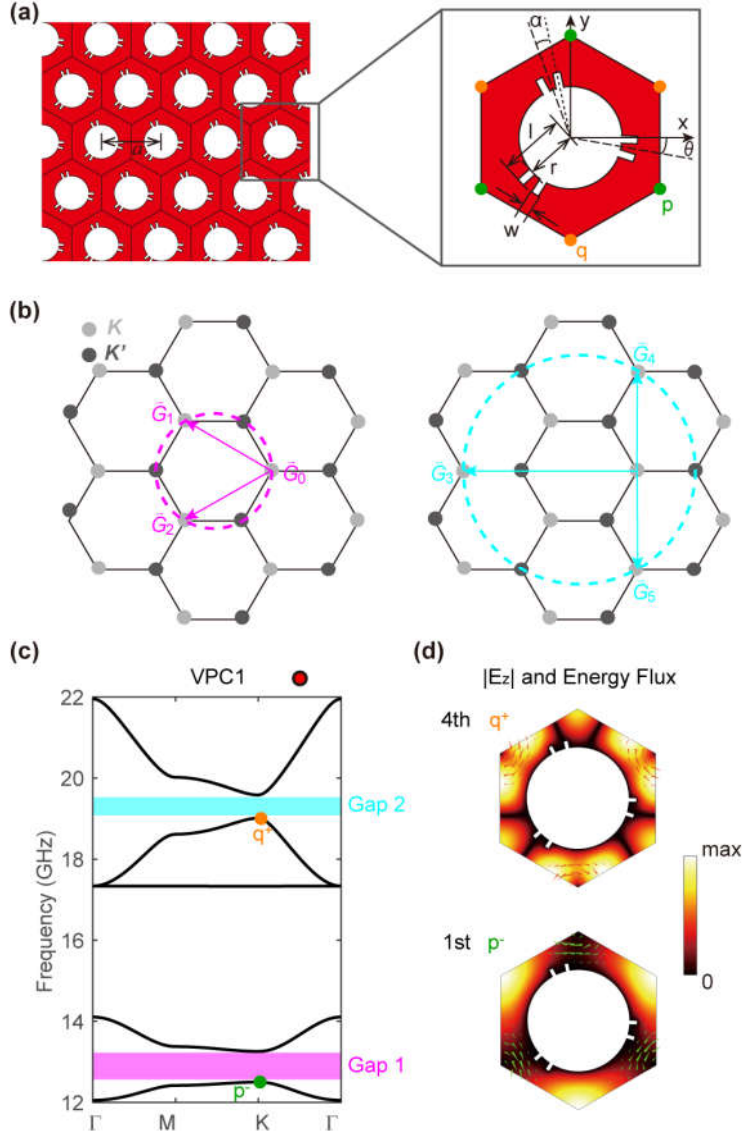

**FIG. 1. Dual band gaps in valley photonic crystals (VPCs).** (**a**) The two-dimensional triangular VPC whose unit cell contains one circle-bar shaped metallic rod in the dielectric background with $\varepsilon$ = 2.25. The lattice constant is given by $a$. The right inset shows five structural parameters of the metallic rod, i.e. ($r$, $l$, $w$, $\theta$, $\alpha$). Labels $p$ and $q$ indicate two different triangular-lattice centers. (**b**) Schematic of two sets of reciprocal lattice vectors used for the plane wave expansion around the K points. The first set makes $|\mathbf{K} + \mathbf{G}_i| = K$ (three magenta arrows at the left) while the second set makes $|\mathbf{K} + \mathbf{G}_i| = 2K$ (three cyan arrows at the right). Within the Brillouin zone, the inequivalent K and K' points are colored in light and dark gray. (**c**) The bulk bands of VPC with $a$ = 20 mm, $r$ = 6.57 mm, $l$ = 7.53 mm, $w$ = 0.5 mm, $\theta$ = -10°, and $\alpha$ = 8° (named as VPC1 for short). Dual band gaps, i.e. Gap 1 and Gap 2 are found. The first (fourth) bulk states at the K point is labelled by the $p^-$ ($q^+$) state according to the position where its $|E_z|$ fields vanish and the winding direction of its energy fluxes. (**d**) $|E_z|$ fields and energy fluxes of the first and fourth bulk states at the K point.

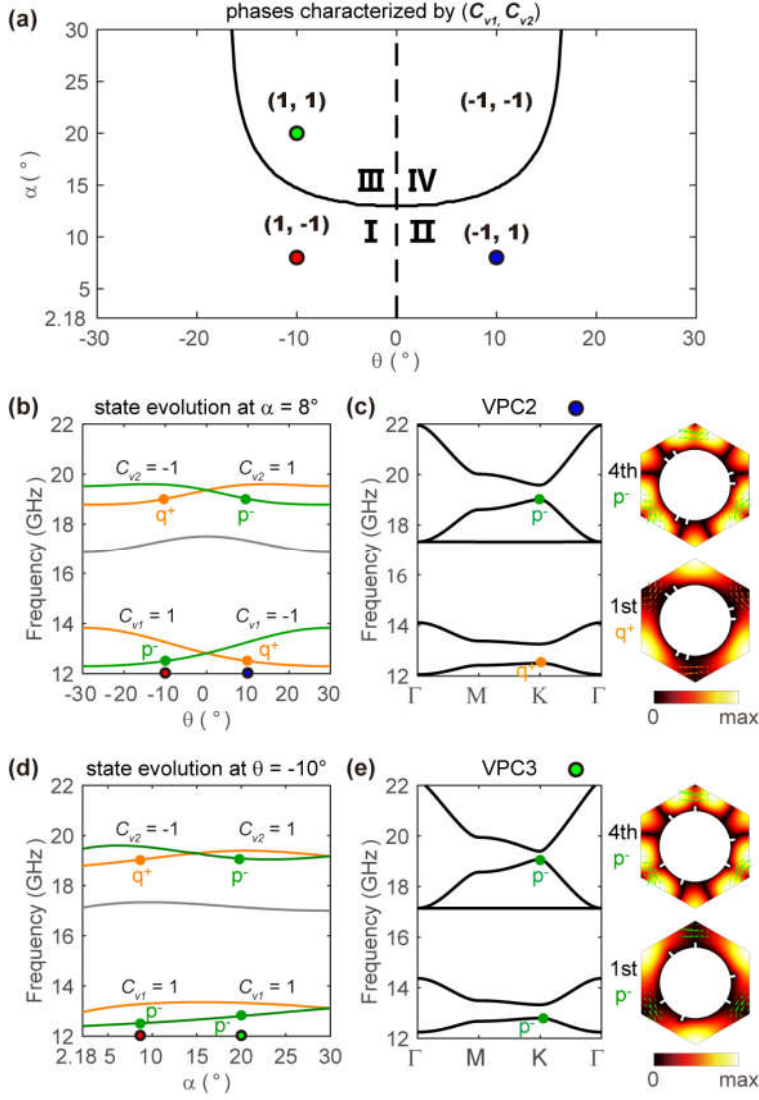

**FIG. 2. Phase diagram and topological phase transition.** (a) The phase diagram of dual band gaps of VPCs with varied $\theta$ and $\alpha$. The dashed (solid) curve show the boundary where the closing of both Gap 1 and Gap 2 (Gap 2) happens. These two curves divide the diagram into four domains which are characterized by a pair of valley Chern numbers ($C_{v1}$, $C_{v2}$) and indexed by the Roman numerals. Three representative VPCs in different phases are labelled by colored dots. (b) The evolution of bulk states at the K point as a function of $\theta$ when $\alpha$ is fixed at $\alpha = 8°$. The phase transition happens at $\theta = 0°$ with the closing of both Gap 1 and Gap 2. (c) The bulk bands of VPC2 with $\theta = 10°$ and $\alpha = 8°$. The $|E_z|$ fields and energy fluxes of the first and fourth bulk states at the K point, i.e. $q^+$ and $p^-$ states, are shown. (d) The evolution of the bulk states at K point as a function of $\alpha$ when $\theta$ is fixed at $\theta = -10°$. The phase transition happens at $\alpha = 14.75°$ with the closing of Gap 2 but not Gap 1. (e) The bulk bands of VPC3 with $\theta = -10°$ and $\alpha = 20°$. The $|E_z|$ fields and energy fluxes of the first and fourth bulk states (both are $p^-$ states) are shown.

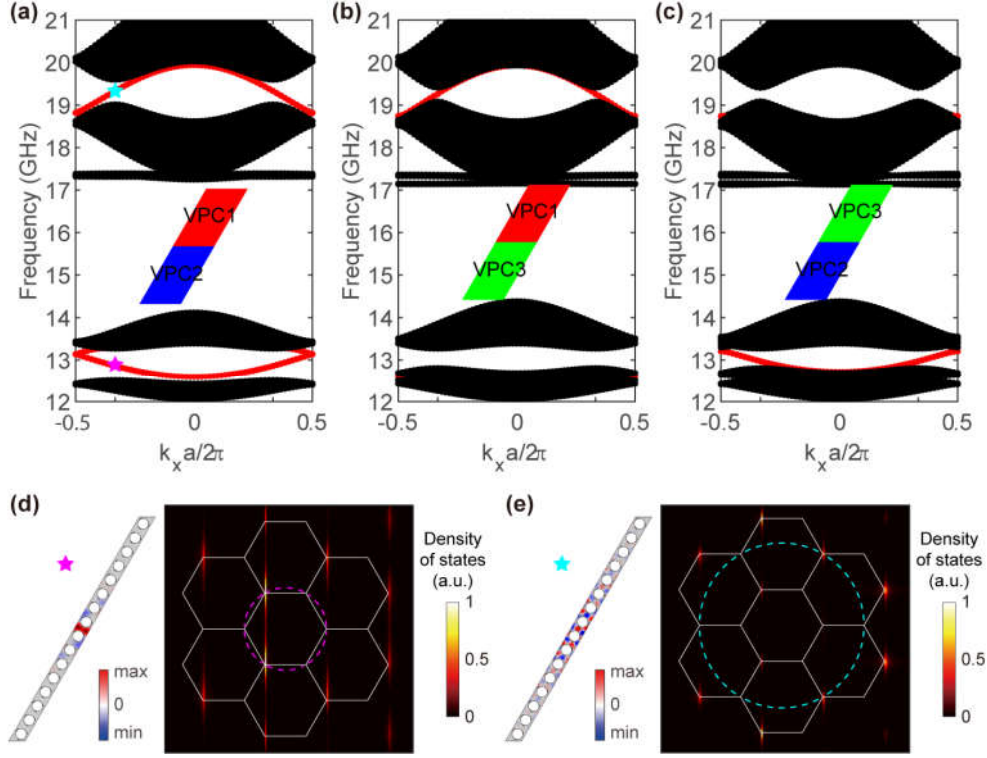

**FIG. 3. Frequency-dependent edge states.** (**a-c**) Edge dispersions of the domain wall between (**a**) VPC1 and VPC2, (**b**) VPC1 and VPC3, (**c**) VPC3 and VPC2. The schematics of domain wall are shown as insets. Valley-dependent edge states are found at both band gaps in (**a**), while they exist in only one band gap in (**b**) and (**c**). (**d**, **e**) The $E_z$ fields and their Fourier transforms of two representative edge states (marked by the cyan and purple stars) of the domain wall in (**a**).

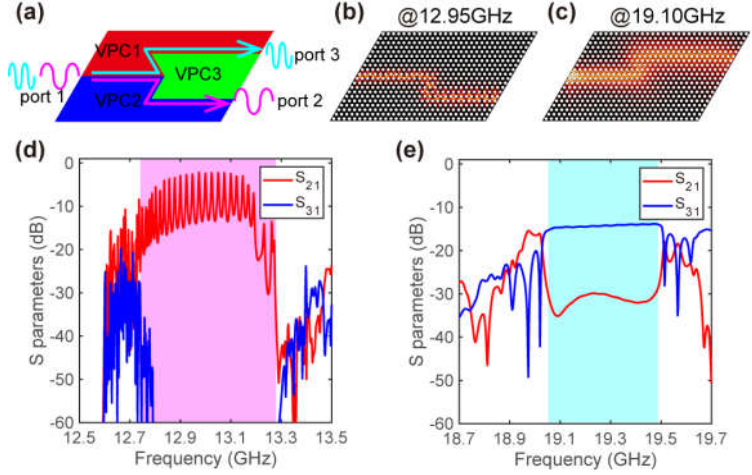

**FIG. 4. Broadband photonic detouring.** (**a**) Schematic of structure for the photonic detouring. The source is input from the left port 1 and received from port 2 or port 3. (**b**, **c**) $|E_z|$ fields of transmitted electromagnetic waves at (**b**) 12.95 GHz and (**c**) 19.10 GHz. (**d**, **e**) The transmission spectra from port 1 to port 2 ($S_{21}$) and from port 1 to port 3 ($S_{31}$) when the source operates with the frequency around (**d**) Gap 1 and (**e**) Gap 2. The transparent boxes highlight the frequency ranges where $S_{21}$ and $S_{31}$ have large difference.

15